Development of a Multi-purpose Fuzzer to Perform Assessment as Input to a Cybersecurity Risk Assessment and Analysis System

Jack Hance
Department of Computer Science
North Dakota State University
1320 Albrecht Blvd., Room 258
Fargo, ND 58108
Email: jack.hance@ndsu.edu

Jeremy Straub
Department of Computer Science
North Dakota State University
1320 Albrecht Blvd., Room 258
Fargo, ND 58108
Phone: +1 (701) 231-8196
Fax: +1 (701) 231-8255
Email: jeremy.straub@ndsu.edu

**Abstract**

Fuzzing is utilized for testing software and systems for cybersecurity risk via the automated adaptation of inputs.  It facilitates the identification of software bugs and misconfigurations that may create vulnerabilities, cause abnormal operations or result in systems' failure.  While many fuzzers have been purpose-developed for testing specific systems, this paper proposes a generalized fuzzer that provides a specific capability for testing software and cyber-physical systems which utilize configuration files. While this fuzzer facilitates the detection of system and software defects and vulnerabilities, it also facilitates the determination of the impact of settings on device operations.  This later capability facilitates the modeling of the devices in a cybersecurity risk assessment and analysis system.  This paper describes and assesses the performance of the proposed fuzzer technology.  It also details how the fuzzer operates as part of the broader cybersecurity risk assessment and analysis system.

**1. Introduction**

The security assessment of critical infrastructure systems is inherently challenging.  A successful penetration test can, in some cases, produce the very malady that the security system seeks to prevent.

For systems that directly support human life and safety and which cannot be taken offline, penetration testing can be too risky to perform. Problematically, due to their criticality, these systems are also the most important to penetration test and protect.

One approach to the security assessment of these critical infrastructure systems is simulation; however, this is inherently constrained by the input data that is utilized. While a digital twin may be possible, in some cases, systems may have hardware components or other characteristics that make real time operational simulation impractical if not impossible. Problematically, this may limit analysis to known system vulnerabilities and known modes of operation – specifically missing the analysis of the 'unknown unknown' that penetration testing can ideally identify.

To facilitate this analysis, a technology that is designed to contribute knowledge to this assessment process is proposed herein. The design for and evaluation of a generic fuzzer designed to support hardware, software and cyber-physical systems that can be readily reused across numerous system types is presented. The fuzzer has a particular focus on systems with configuration files and seeks to understand the impact of different values, in these files, on system operations, vulnerabilities and resilience. This data collection capability facilitates modeling these systems, in a simulation system, in a manner that is indicative of how they actually work – instead of how they are supposed to.

This paper presents and evaluates the proposed fuzzer system. It continues with a discussion of prior work, in Section 2. Section 3 presents the system's design. Then, in Section 4, system operations are discussed and an example of the system's operations is presented in Section 5. Section 6 presents data and analysis regarding the evaluation of system performance and efficacy, before the paper concludes and discusses prospective topics for future work in Section 7.

**2. Background**

This section presents prior work that informs the work that is presented herein. First, prior work on the penetration testing of critical infrastructure systems is discussed. Then, work on fuzzing technologies is reviewed.

*2.1. Penetration Testing of Critical Infrastructure Systems*

Due to their criticality and growing interconnectedness [1], a significant focus has been placed on techniques and technologies to secure and evaluate the security of critical infrastructure and SCADA systems. Prior work in this area has taken several forms.

Some have developed [2,3] or proposed [1] test labs and platforms. Ficco, Choras and Kozik [4] proposed a "hybrid and distributed simulation platform" that combined the ability to incorporate actual hardware and systems in some areas while simulating other areas and used a cloud architecture.

An alternate approach was proposed by Turpe and Eichler [5] who suggested conducting penetration testing on live production systems while incorporating precautions and risk mitigation methods. These would restrict tests to a "selection of test cases and techniques" and use isolated subsystems. Problematically, this would preclude testing some scenarios which may be the most catastrophic, if they eventuated in a real attack, to avoid risks during the testing.

Ralethe demonstrated testing and risk categorization using a "virtual plant environment" [6]. Rocha [7], similarly, used a simulation environment (the EFACEC ScateX# SCADA system) to demonstrate a methodology for SCADA system security assessment. This approach utilized standards assessment and a threat model, which identified entry points and targets and utilized attack vector characterization, to identify risk factors.

Knowles, Baron and McGarr [8] proposed the standardization of penetration testing, which would be of particular importance for critical infrastructure-level systems, and introduced a framework to serve this purpose. Speicher [9] proposed a technique called "Stackelberg planning" that utilizes a network model which pits a simulated attacker against defender actions. Li, Yan and Naili [10] have proposed the use of artificial intelligence, in the form of deep reinforcement learning, to find the "optimal attack path against system stability".

A variety of frameworks have also been proposed for internet of things (IoT) devices [11], which are informative to the challenge of penetration testing critical infrastructure systems. Several simulated IoT devices of various types. Two, though, were of particular interest to this work. Brown, Saha and Jha [11,12] proposed a framework that focused on the impact of an attack on the entire system or network, instead of just on whether an attack would be successful, and Lee, et al. [11,13] proposed an approach that incorporated a fuzzing component that was targeted at software defined networks.

*2.2. Fuzzing Technology*

Fuzzing software manipulates system inputs to assess their impact on a system's performance. A variety of types of fuzzing software have been previously developed.

In many cases, fuzzers are application-specific and not portable without code modifications. Ispoglou, et al. [14], though, developed a fuzzer that automatically assesses libraries using an entirely automated analysis process.

Fuzzers have also historically not dealt with system state [15]; however, Banks, et al.'s SNOOZE fuzzer introduced state awareness and was used to test the voice over internet protocol (VoIP) session initiation protocol (SIP). Pham, Bohme and Roychoudhury [16] also considered state in their AFLNet fuzzer, which is designed to automate fuzzing by watching, learning from, manipulating and then replaying traffic between a client and server. Li, et al. [17], similarly, developed a 'grey box' (mostly opaque) fuzzer, called SNPSFuzzer, that was shown to increase coverage and processing speed, as compared to the AFLNet fuzzer.

While many fuzzers are 'black boxes' (opaque) or 'grey boxes', Jayaraman, et al. [18] introduced a 'white box' (transparent) fuzzer that utilizes code-based knowledge to ensure that it fully tests all possible program paths. Another advancement was made by Lee, et al. [19] who developed a command line option-aware fuzzer that was shown to outperform many modern fuzzers due to its enhanced searching capability. Chen, et al. proposed a 'grey box' tester (called Hawkeye) which uses static analysis and can focus on user-identified program targets [20].

Zou, et al. [21] built upon existing fuzzers by developing a system called SyzScope. This software was designed to analyze existing known low-risk vulnerabilities to assess if they provided higher-than-anticipated risk levels or access to a higher priority exploit.

Riterau-Brostean, et al. [22] built a fuzzer that develops a model of clients and servers which use the DTLS protocol. It utilizes "model learning" to develop a state machine model which can then be tested or analyzed to identify prospective security vulnerabilities [22].

Supporting contributions were also made by Lipp, et al. [23], who created a coverage analyzer for fuzzer testing, and Metzman, et al. [24], who developed an "open fuzzer benchmarking platform and service". Metzman, et al. [24] note that "proper evaluation of fuzzing techniques remains elusive", while hoping that their benchmarking platform and service can aid in fuzzer evaluations.

**3. System Design**

This section presents the design of the proposed fuzzer system. First, the system's goals are presented. Then, a summary of its design is presented. Following this, the individual components are discussed. Finally, a summary of the execution flow is provided.

*3.1. System Goals*

The goal of the proposed fuzzer system is to provide a framework that allows the testing of the effects of different configurations on any target system. A system such as this requires a large amount of flexibility and may have performance limitations, as compared to single-purpose fuzzers, in order to make it as widely usable as possible. Uses of this system include fuzzing configs for program bugs and errors, finding the minimum changes that are needed to put a system in a desired state, and finding configs for which a given exploit is effective.

The proposed system is designed to support the fuzzing of a wide variety of targets. Each target can have its configuration values changes and then be scanned and have its new state logged. An example use of a system such as this is to characterize a generic managed switch device, where the functionality of running custom programs on the system is restricted and changes to configuration values can have outward-facing effects. Fuzzing a target such as this could reveal bugs related to configuration changes and system vulnerabilities.

*3.2. Summary of Design*

The proposed system, an overview of which is shown in Figure 1, consists of five distinct parts: the server, the client, the communicator, the config definition, and the results database. The server commands the system. It generates config changes, pushes them to the target, executes tests against the target, and then stores the results. The client acts as an interface between the communicator and the server, pushing config change requests from the server to the communicators and passing back status updates from the communicator.

The communicator is a component that is custom developed for each target and is used to push config changes and (optionally) check the validity of each new config. The config definition stores a description of the config being enumerated with a description for each configurable value. These are described in a way that allows new values to be procedurally generated by the server. The results database stores each config change and the related test results.

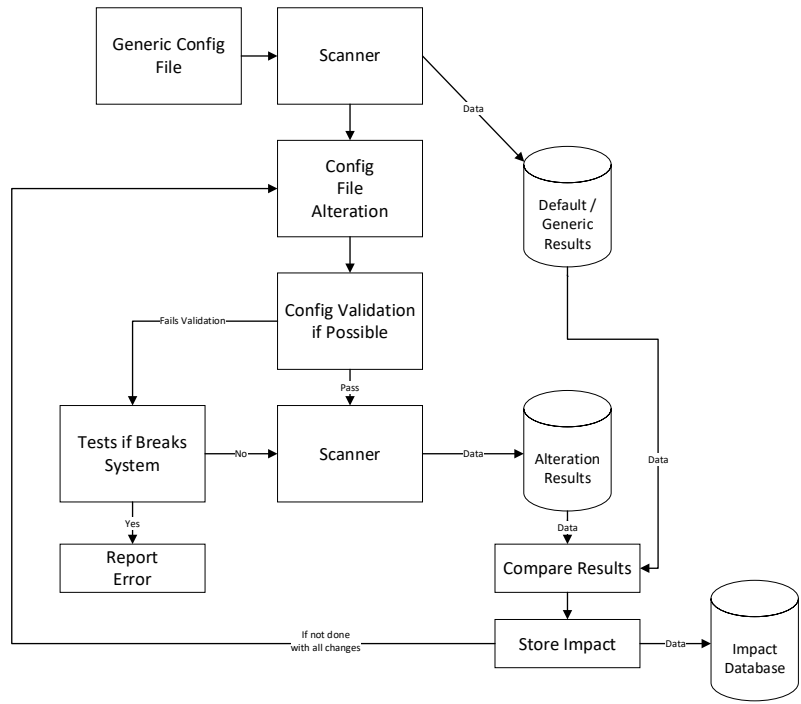

**Figure 1.** System Overview.

## 3.3. Description of Components

This section describes the components that are part of the proposed fuzzer system. A conceptual overview of the system is shown in Figure 2. The presentation of the components begins with the server, which is discussed in Section 3.3.1. Next, the client is described in Section 3.3.2 and the communicator is described in Section 3.3.3. Finally, the config definition and results database are described in Sections 3.3.4 and 3.3.5, respectively.

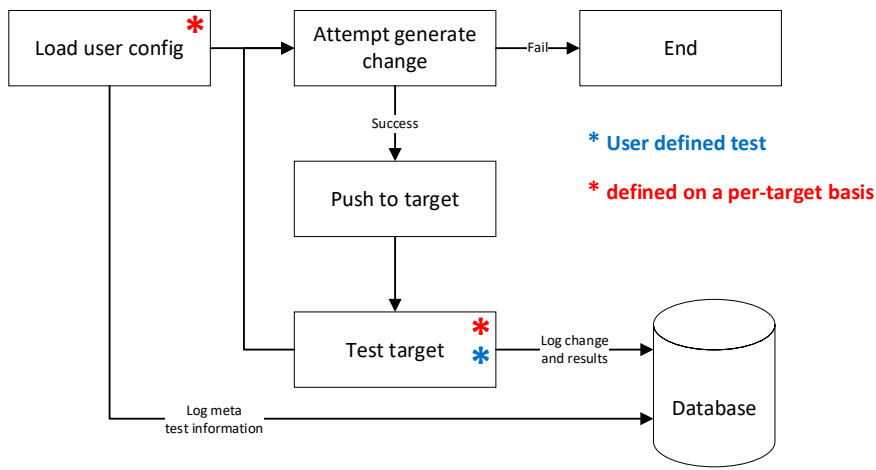

Figure 2. Conceptual overview of system.

*3.3.1. Server*

The server commands the fuzzing system. It has four major components: configuration description parsing, config change generation, testing, and logging. Configuration parsing involves reading and parsing a configuration description, which describes how values are generated for config change generation. Config change generation takes the described values from the configuration description and sequentially generates new changes to be pushed to the target. After each config change is pushed, a set of tests are executed against the target to attempt to detect a state change. Logging simply takes the test results and config changes and stores them for future analysis.

Config changes are derived from the config definition. Their definition is parsed and new values are generated as needed. How config changes are defined and generated is described in more detail, subsequently. These config changes are then passed to the client to be pushed to the communicator. As the client is very likely to be running on the target machine, the sever that communicates these changes over the network is hosted on the server. Opening a port on the server causes less interference with the network fingerprint of the target. However, it requires a reverse connection mechanism where the server knows when it has new data, but the client needs to initialize the connection to retrieve it.

Tests executed by the server are customizable on a target-by-target basis, and tests can be custom implemented to suit a target's testing and customization needs. At present, tests are executed at a network level, but as the current implementation has the client and communicator running on the same machine as the target, tests could also be executed via the client to allow for more in-depth testing. The only non-service-specific network level test, at present, is a generic port scan. The system provides the ability to add new tests, which allows service specific tests to be added, if applicable to the target. An example of a service specific scan could be something such as directory and file enumeration for an HTTP or FTP service.

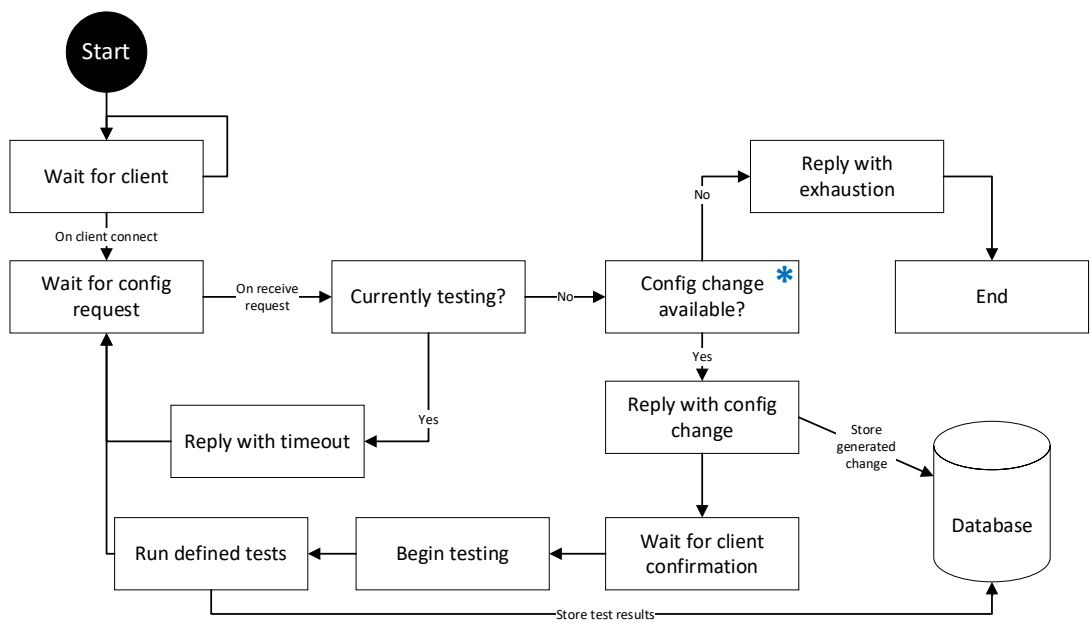

\* **Queries change generator**

**Figure 3.** Server functionality.

Tests are implemented in Python and use a provided library to make communication with the server more convenient. Tests are called with a set of parameters that describe the state and location of the target. They end by returning their results in a text-based form to be stored in the result database. Tests are called as subprocesses of the server, and communications are performed via stdin and stdout.

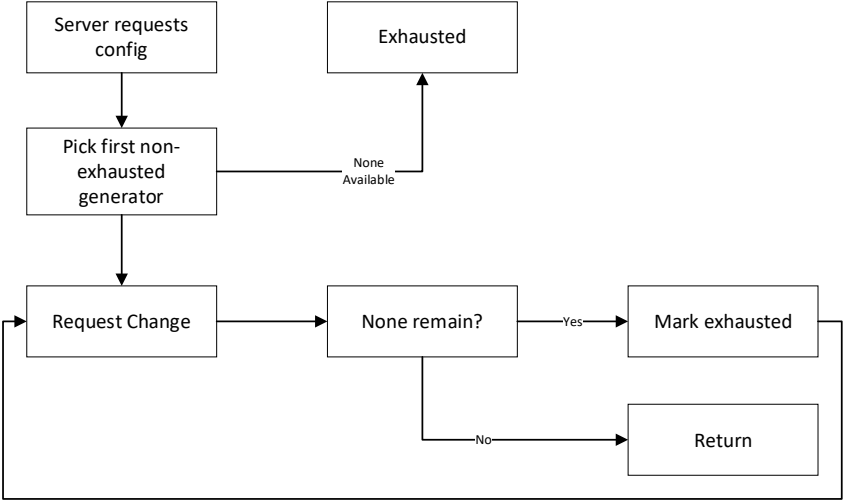

Figure 4. Change Generator – Generation.

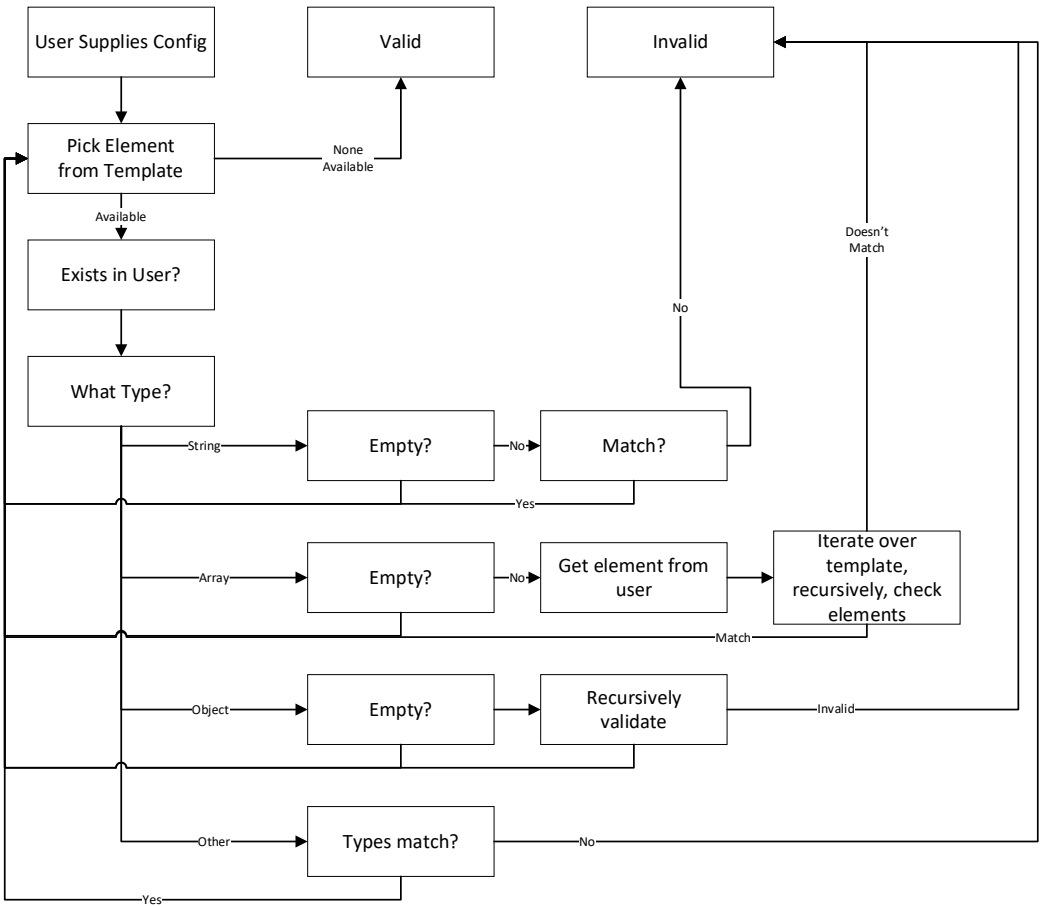

Figure 5. Change Generator – Validation.

For each new target, a configuration description has to be created. The configuration description describes the config parameters to be changed and what values to change them to. At present there are three different parameters types: strings, numbers, and Booleans. Strings can have discrete values passed (i.e., ["stringA", "stringB", "stringC"]).  Strings can also be generated using regular expressions. Numbers can also have discrete values or they can be defined in terms of ranges of numbers. Booleans have the vale of either true or false. All parameters also provide a default value to reset the parameter back to after it has been tested.

The configuration description is parsed and converted into a set of parameter generators.  Each parameter generator generates a set of values as described in the configuration description. At present, the set of parameter generators is processed sequentially: each value is passed, in sequence, from the first parameter, then it is reset, and then the second, and so on until all values have been exhausted. Future versions of the system will allow different configuration values to be tested at the same time.  It is also planned to generate configuration values in an intelligent way that is designed to seek out the desired results.

After a configuration change has been passed and implemented by the communicator (which will be discussed in more depth shortly), the server then runs a set of tests against the target to determine its state.

After the config change has been passed and the target has been tested, the change and the tests results are stored for future analysis in a SQLite database.

*3.3.2. The Client*

The client operates as a network interface between the communicator and the server. Its functionality is depicted in Figure 6. All of the current functionality of the fuzzer *could* be implemented within the communicator, but as a new communicator has to be implemented for *every* new target it is more efficient to place the common functionality in a separate executable. The client accepts a basic configuration file that describes the communicator and the IP address and port of the server. The communicator is described in further depth in the next section. It is implemented in Python and uses custom library to make communications with the client hassle-free and standardized. The client starts the Python script as a sub-process and then communicates with it using stdin and stdout. Beyond this the client simply acts as a proxy between the communicator and the server, with no other processing being performed on transmissions between them.

An added benefit of the client and server being separate processes that communicate over the network is that it limits the footprint needed on the target computer. It also allows the target to be fully restarted without impacting the server process.

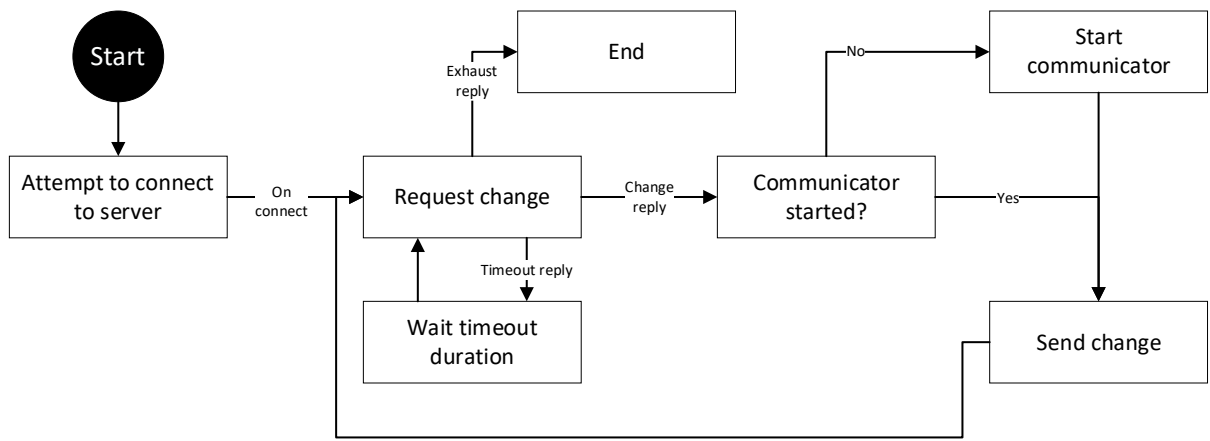

**Figure 6.** Client functionality diagram.

*3.3.3. The Communicator*

The communicator is a Python script that is custom created for each new target. Its functionality is Depicted in Figure 7. It allows config changes to be pushed and (optionally) validated to the target. The reason for having the communicator implemented this is was to allow for the system to interface with as many targets as possible. For example, the way Apache2 (an HTTP server) is configured is vastly different from how a managed ethernet switch (that uses serial for configuration) is configured. Using custom implemented communicators allows the fuzzer system to be used with both of these targets.

The communicator is strongly tied to the implementation of the config database (which is discussed shortly). Generated values from the config database can be interpreted and pushed differently at the discretion of the communicator, as it is fully scriptable and independent from the rest of the system.

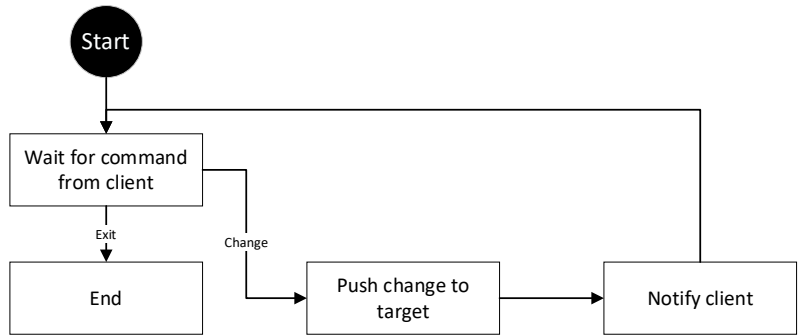

**Figure 7.** Communicator functionality diagram.

The Python library for the communicator is designed in such a way that the communicator can operate in either a stateful or stateless mode, at the discretion of the implementor.

The communicator's role is to take config changes from the server and push them to the target. A future planned enhancement is having the communicator also validate that the config change was accepted by the target. The communicator is a python script that is custom implemented on a target-by-target basis, providing the ability to fuzz as many different devices as possible. Some different examples of what the communicator could be implemented for are:

- A local server that accepts config changes via a configuration file
- A firewall that accepts changes via a web interface
- A managed switch that accepts changes via a serial connection.

By having the communicator implemented in Python, as long as configuration values can be pushed programmatically, any software or device it can communicate with can be fuzzed using the proposed system.

*3.3.4. The Config Definition*

The config definition stores definitions. It houses a standardized definition of the config of the target machine, with each configurable value defined in a way that allows for new values to be generated by the server. The config definition does not need to exactly represent the config of the target, as the communicator decides how values are pushed to the target. As long as the communicator understands the parameter and the provided value, how it is pushed to the target is independent of the rest of the system. For example, a router that might be configured via a graphical web interface, and typical text values may not make as much sense as abstract values that are interpreted by the communicator.

Configs are described in JSON, with each value being assigned a type and having either a generating expression or a discrete list of values. Support for additional types is planned in the future.

Config changes are defined by a key, an action, and action parameters. Example actions are modify, add, delete, and reset. Modify is the most common command, it takes an existing key and changes its value. Add adds a new value, which is a valid config value but which may not already be set or exist in the target system. Delete removes a value from the config (e.g., by commenting out or removing a line in a config file). Reset sets a value back to its default, though how this is performed is determined by the communicator.

*3.3.5. The Result Database*

The result database stores the changed key or keys, their new values, and the results of the tests that have been executed with these config values. The result database is implemented in SQLite to allow easy offline use while still maintaining the ability to quicky analyze the data after testing.

*3.4. Summary of Execution Flow*

After the client and server have started and have connected, the execution of the system proceeds as follows:

1. The server generates a new config change.
2. The server passes this config change to the client.
3. The client passes this config change to the communicator.
4. The communicator then applies this change to the target.
5. (Optional) The communicator uses a built-in mechanism to check if the config is valid.
6. The communicator reports back to the client that the config change was applied and, if its validity was tested, the results of the validity test
7. The client passes the config change status to the server.
8. The change and the test results are stored in the results database.
9. If all config changes have been exhausted, execution ends.  If there are config changes left, then return to step 1.

**4. System Operations**

This section presents the operations of the fuzzing system.  First, an overview of operations is provided. Then, the system operations are discussed in the context of the cybersecurity simulation assessment system.

*4.1. Operations Overview*

The operations of the proposed system are complex, so a high-level description is provided:

*Server Start*

The server is provided a port number to bind to and a configuration description. The description is validated and, if meets specifications, then the server binds to the port and awaits a client connection.

*Client Start*

The client is provided the IP and port of the server along with the path to the communicator python script. The client attempts to connect to the server using the provided parameters.

*Handshaking*

Upon successful connection to the server, the client sends a HANDSHAKE_INIT request. The server receives this and replies with a HANDSHAKE_ACK response. After sending the response, the server transitions to the config change request state.  The client does the same, upon receiving the response.

*Config Request*

The client sends the server a CONFIG_REQUEST message.  The server has three different response types to send in response to this request:
- If a config change is available and the server is not currently executing tests, a CONFIG_FULFILLMENT message is returned.
- If the server is currently testing, a CONFIG_TIMEOUT message is returned.
- If the server has used all available config changes, a CONFIG_EXHAUSTION message is returned.

*Server response: CONFIG_FULFILLMENT*

The server attempts to generate a change, if a change is successfully generated then the parameter name and new value are sent to the client. If the communicator has not yet been started by the client, the communicator is started. The client then forwards the change to the communicator and replies to the server with a CONFIG_CONFIRMATION response. Upon receiving a CONFIG_CONFIRMATION response, the server transitions to a testing state and immediately starts running tests against the target.

*Communicator Config Change Application*

Once the communicator receives a config change, it applies this to the target. How the change is applied to the target is specific to the communicator implementation and it is abstracted from the Client. Once the communicator has applied the change, it indicates this to the client and it then waits for the next change.

*Testing*

After a CONFIG_CONFIRMATION is received by the client, testing is started against the target device.  Modular tests are created on a per-target basis.  The tests are also implemented in python and selected for use in the configuration description. The server will return CONFIG_TIMEOUT responses during the entire time testing is executed.  Once it is finished, the server returns to the config request state.

*Logging*

The config change along with its tests' results are stored in the results database when testing is finished.

*Server response CONFIG_TIMEOUT*

If the server replies with a CONFIG_TIMEOUT response, the client waits for the time specified in the response and then sends another CONFIG_REQUEST message.

*Server response CONFIG_EXHAUSTION*

If the server has exhausted the config changes it has available, it returns a CONFIG_EXHAUSTION message.  Upon the client receiving this response, it tells the communicator to disconnect from the target and stop execution. After the communicator has closed, the client also closes. The server also stops its execution.

*4.2. Operations in Context*

The data collected by the fuzzing system can serve as an input to the identification of vulnerability paths through the network of the IT system which is being assessed. This facilitates analysis regarding what the impact of each potential setting value or settings change would be on the network and the computers and other devices connected to it. Figure 8 shows how the configuration parameter information (i.e., information about what parameter values are valid) is provided to the fuzzing scanner, which conducts a scan of the target device. Then, the inputs and results for each parameter value combination are stored in a settings and impact database. This database, along with knowledge of the interconnectivity of the computing system and data regarding known vulnerabilities, in general, can serve as inputs to a vulnerability path generation mechanism. This can help staff to understand the overall security of the IT system and, based on the frequency of encountering a vulnerability, it can assist cybersecurity staff in prioritizing vulnerable systems for attention.

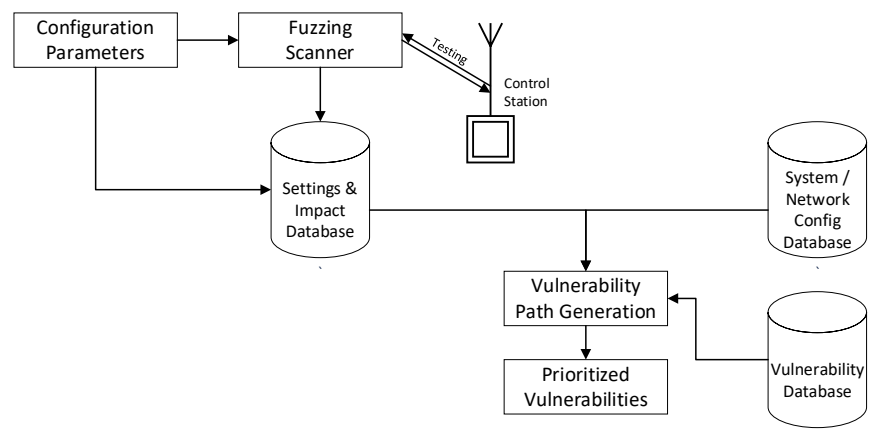

**Figure 8.** Operations in context.

## 5. Example of Operations

A basic example system is now discussed to demonstrate the fuzzer's operations. It is implemented as follows:
- The target device is a virtual machine running Ubuntu 22.04.1 LTS, with UFW used as a firewall
- The configuration description describes a range of port numbers
- Config changes are made by issuing commands to UFW

The client is run on the Ubuntu virtual machine, and the server is run on a separate Linux virtual machine. The two machines are networked. The setup models a system running software that can have the port it binds to be changed by configuration changes.

The full configuration description used for this example is shown in Listing 1.

**Listing 1.** Configuration example.

```json
{
    "meta": {},
    "parameters": [
        {
            // ports to open on ufw
            "pname": "port",
            "ptype": "number",
            "pdefault": "4999",
            "pvalues": [
                {
                    "value_type": "range",
                    "value": {
                        "start": 5000,
                        "end": 5010
                    }
                }
            ]
        }
    ]
}
```

The configuration description has a parameter named "port," and it targets a range of port values from 5000 to 5009. A default value of 4999 is provided.

A communicator script was also created for this example. It is presented in Listing 2.:

**Listing 2.** Communicator script example.

```python
from lib import *
import subprocess

class ExampleCommunicator(TargetCommunicator):
    # ready method is called right after run is called, this can be used
    # to create any stateful variables
    def ready(self):
        self._enabled_fw = False

    def close(self):
        if self._enabled_fw:
            # reset the firewall on exit
            subprocess.run(["ufw", "--force", "reset"], capture_output=True)

    # on_command is called whenever a new config_request is received, it
    # requires that a CommunicatorResponse object is returned
    def on_command(self, config_request: ConfigRequest):
        if config_request.command.upper() == "CHANGE":

            if not self._enabled_fw:
                subprocess.run(["ufw", "enable"], capture_output=True)
                self._enabled_fw = True

            c_name = config_request.config_change["Name"]
            c_val = config_request.config_change["Value"]
```

```python
            if c_name.upper() == "PORT":
                # ufw is reset to clean out prior changes
                subprocess.run(["ufw", "--force", "reset"], capture_output=True)
                # ufw is re-enabled (ufw is disabled on reset by default)
                subprocess.run(["ufw", "enable"], capture_output=True)
                # the new port value is pushed
                subprocess.run(["ufw", "allow", f"{c_val}"], capture_output=True)
                # netcat is started on the port to model a running service, as otherwise port scanners are unable to
                # detect the change
                subprocess.Popen(["nc", "-lvp", f"{c_val}"], stdout=subprocess.DEVNULL, stderr=subprocess.DEVNULL)

        return CommunicatorResponse(
            status = ConfigChangeStatus.OK,
            extended_status = {}
        )

if __name__ == "__main__":
    comm = ExampleCommunicator()
    comm.run()
```

For this example, the communicator opens a port in UFW and netcat is started to simulate a server process, as otherwise port scanners would be unable to detect the change.

The server consumes the configuration description given above, and then waits for the client to connect to it. Once the client has connected, configuration changes are pushed and tests are executed until the settings values in the configuration description are exhausted.

Figure 9 shows an image of the server's output.

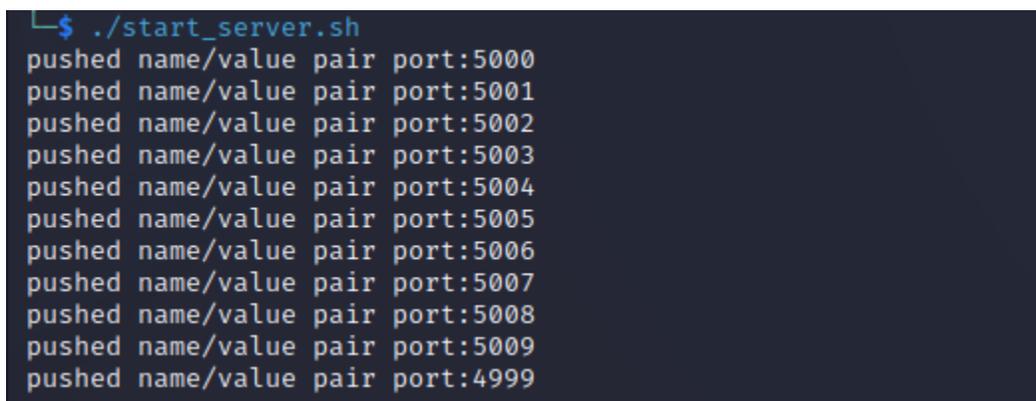

**Figure 9.** Example output.

The client does not generate any output. The server generates a CSV, which is shown in Figure 10.

| | A | B | C |
|---|---|---|---|
| 1 | changeNa | "changeR | "ports_open" |
| 2 | port | 5000 | 5000 |
| 3 | port | 5001 | 5001 |
| 4 | port | 5002 | 5002 |
| 5 | port | 5003 | 5003 |
| 6 | port | 5004 | 5004 |
| 7 | port | 5005 | 5005 |
| 8 | port | 5006 | 5006 |
| 9 | port | 5007 | 5007 |
| 10 | port | 5008 | 5008 |
| 11 | port | 5009 | 5009 |
| 12 | port | 4999 | 4999 |

**Figure 10.** Server output CSV.

In Figure 10, the first column names the parameter that was changed ("port"), the second column lists the value it was changed to (the new port number), and the last column lists the ports that were detected as open (in this case only the port that was sent was open, so the third column mirrors the second). This simple event serves to demonstrate the basic functionality of the fuzzer system as well as to demonstrate the basic process that will be used for more complex fuzzing tasks.

## 6. Experimentation, Data and Analysis

This section discusses the experimental validation of the proposed fuzzing technology and system. In Section 6.1, the experimental process is described. Then, in Section 6.2, data is presented and analyzed.

### 6.1. Experimental Process

Presently, the fuzzer is designed to test a single value change at a time (fuzzing multiple values can be accomplished through multiple runs and functionality to expedite multiple-value fuzzing is planned to be added in the future). Given this, the current process is as follows:

- Generate a change from one of the generator definitions in the main config file
- Pass the change to the communicator, which applies it to the target
- Run all tests
- Once all changes have been implemented for this generator, reset the target to the specified default value
- Repeat step 1 until all generators have been exhausted

The target that was used for this testing was Apache2, a popular HTTP server. The methodology used was to implement multiple config changes that yield observable state changes, and then to implement a test for each one of these. The config changes, which are discussed in more detail in Appendix A, that were implemented were:

- Enabling and disabling the server entirely
- Changing which port the server binds to
- Changing how the server version is returned in the header of responses
- Changing how the server version is returned at the bottom of error pages

Tests, which are presented in more detail in Appendix B, were implemented to test for each of these bullet points, along with two additional tests that are not affected by implemented config changes:

- Testing for CVE-2021-41773
- Testing for CVE-2021-42013

Most of the tests were designed to validate the projected results. The two CVE detection tests were added to demonstrate how the fuzzer could detect if unexpected states are achieved by changing apparently unrelated values. There was no specific reason to believe that CVE-2021-41773 or CVE-2021-42013 would be enabled by the configuration changes that were made; however, numerous tests of this type could be included looking for known issues and issue types that may be produced by various configuration changes.

In this case, the value changes did not reenable the CVE-2021-41773 or CVE-2021-42013 vulnerabilities.

### 6.2. Data and Analysis

This section presents the data that was collected using the procedures described in Section 6.1. Table 1 presents the fuzzer test instructions which were generated to collect the data using the described procedures. As shown in the table, test IDs 1 to 3 alter the value of the start_systemctl_service, test IDs 4 to 104 change the value of the port setting, test IDs 105 to 108 change the value of the server signature setting and test IDs 109 to 115 change the value of the server tolkens setting. Notably, these are all settings within the Apache server configuration file that have been identified based on the settings file format definitions.

**Table 1.** Fuzzing test instructions (note that brackets indicate summarized values).

| Id | ChangeName | ChangeValue |
|---|---|---|
| 1 | start_systemctl_service | FALSE |
| 2 | start_systemctl_service | TRUE |
| 3 | start_systemctl_service | TRUE |
| [4-103] | port | [30000-30099] |
| 104 | port | 80 |
| 105 | server_signature | On |
| 106 | server_signature | Off |
| 107 | server_signature | EMail |
| 108 | server_signature | On |
| 109 | server_tokens | Full |
| 110 | server_tokens | Prod |
| 111 | server_tokens | Major |
| 112 | server_tokens | Minor |
| 113 | server_tokens | Min |
| 114 | server_tokens | OS |
| 115 | server_tokens | OS |

Using the settings values indicated in Table 1, data was collected using the five different data collection mechanisms, which were previously described. Table 2 presents the data that was collected for the header version returned. Table 3 presents the data that was collected for the page version returned. Tables 4 and 5 present data regarding about whether exploitation of the CVE-2021-41773 and CVE-2021-42013 vulnerabilities (respectively) was detected. Finally, Table 6 presents the results of a port scan.

**Table 2.** Header version results (note that brackets indicate summarized data).

| ChangeId | ResultName | ResultSummary |
|---|---|---|
| 1 | header version | <could not find server> |
| [2-109] | header version | Apache/2.4.53 (Debian) |
| 110 | header version | Apache |
| 111 | header version | Apache/2 |
| 112 | header version | Apache/2.4 |
| 113 | header version | Apache/2.4.53 |
| 114 | header version | Apache/2.4.53 (Debian) |
| 115 | header version | Apache/2.4.53 (Debian) |

**Table 3.** Page version results (note that brackets indicate summarized data).

| ChangeId | ResultName | ResultSummary |
|---|---|---|
| 1 | page version | <could not find server> |
| 2 | page version | Apache/2.4.53 (Debian) Server at 127.0.1.1 Port 80 |
| 3 | page version | Apache/2.4.53 (Debian) Server at 127.0.1.1 Port 80 |
| [4-103] | page version | Apache/2.4.53 (Debian) Server at 127.0.0.1 Port [30000-30099] |
| 104 | page version | Apache/2.4.53 (Debian) Server at 127.0.0.1 Port 80 |
| 105 | page version | Apache/2.4.53 (Debian) Server at 127.0.0.1 Port 80 |
| 106 | page version | <no version number found> |
| 107 | page version | Apache/2.4.53 (Debian) Server at <a href="mailto:webmaster@localhost">127.0.0.1</a> Port 80 |
| 108 | page version | Apache/2.4.53 (Debian) Server at 127.0.0.1 Port 80 |
| 109 | page version | Apache/2.4.53 (Debian) Server at 127.0.0.1 Port 80 |
| 110 | page version | Apache Server at 127.0.0.1 Port 80 |
| 111 | page version | Apache/2 Server at 127.0.0.1 Port 80 |
| 112 | page version | Apache/2.4 Server at 127.0.0.1 Port 80 |
| 113 | page version | Apache/2.4.53 Server at 127.0.1.1 Port 80 |
| 114 | page version | Apache/2.4.53 (Debian) Server at 127.0.1.1 Port 80 |
| 115 | page version | Apache/2.4.53 (Debian) Server at 127.0.1.1 Port 80 |

**Table 4.** CVE-2021-41773 results (note that brackets indicate summarized data).

| ChangeId | ResultName | ResultSummary |
|---|---|---|
| 1 | CVE-2021-41773 | <could not find server> |
| [2-115] | CVE-2021-41773 | target is not vulnerable |

**Table 5.** CVE-2021-42013 results (note that brackets indicate summarized data).

| ChangeId | ResultName | ResultSummary |
|---|---|---|
| 1 | CVE-2021-42013 | <could not find server> |
| [2-115] | CVE-2021-42013 | target is not vulnerable |

**Table 6.** Port scan results (note that brackets indicate summarized data).

| ChangeId | ResultName | ResultSummary |
|---|---|---|
| 1 | port scan | [(30000, False), (30001, False), (30002, False), (30003, False), (30004, False), (30005, False), (30006, False), (30007, False), (30008, False), (30009, False), (30010, False), (30011, False), (30012, False), (30013, False), (30014, False), (30015, False), (30016, False), (30017, False), (30018, False), (30019, False), (30020, False), (30021, False), (30022, False), (30023, False), (30024, False), (30025, False), (30026, False), (30027, False), (30028, False), (30029, False), (30030, False), (30031, False), (30032, False), (30033, False), (30034, False), (30035, False), (30036, False), (30037, False), (30038, False), (30039, False), (30040, False), (30041, False), (30042, False), (30043, False), (30044, False), (30045, False), (30046, False), (30047, False), (30048, False), (30049, False), (30050, False), (30051, False), (30052, False), (30053, False), (30054, False), (30055, False), (30056, False), (30057, False), (30058, False), (30059, False), (30060, False), (30061, False), (30062, False), (30063, False), (30064, False), (30065, False), (30066, False), (30067, False), (30068, False), (30069, False), (30070, False), (30071, False), (30072, False), (30073, False), (30074, False), (30075, False), (30076, False), (30077, False), (30078, False), (30079, False), (30080, False), (30081, False), (30082, False), (30083, False), (30084, False), (30085, False), (30086, False), (30087, False), (30088, False), (30089, False), (30090, False), (30091, False), (30092, False), (30093, False), (30094, False), (30095, False), (30096, False), (30097, False), (30098, False), **(30099, False)**]<br>… |
| 103 | port scan | [(30000, False), (30001, False), (30002, False), (30003, False), (30004, False), (30005, False), (30006, False), (30007, False), (30008, False), (30009, False), (30010, False), (30011, False), (30012, False), (30013, False), (30014, False), (30015, False), (30016, False), (30017, False), (30018, False), (30019, False), (30020, False), (30021, False), (30022, False), (30023, False), (30024, False), (30025, False), (30026, False), (30027, False), (30028, False), (30029, False), (30030, False), (30031, False), (30032, False), (30033, False), (30034, False), (30035, False), (30036, False), (30037, False), (30038, False), (30039, False), (30040, False), (30041, False), (30042, False), (30043, False), (30044, False), (30045, False), (30046, False), (30047, False), (30048, False), (30049, False), (30050, False), (30051, False), (30052, False), (30053, False), (30054, False), (30055, False), (30056, False), (30057, False), (30058, False), (30059, False), (30060, False), (30061, False), (30062, False), (30063, False), (30064, False), (30065, False), (30066, False), (30067, False), (30068, False), (30069, False), (30070, False), (30071, False), (30072, False), (30073, False), (30074, False), (30075, False), (30076, False), (30077, False), (30078, False), (30079, False), (30080, False), (30081, False), (30082, False), (30083, False), (30084, False), (30085, False), (30086, False), (30087, False), (30088, False), (30089, False), |

(30090, False), (30091, False), (30092, False), (30093, False), (30094, False), (30095, False), (30096, False), (30097, False), (30098, False), **(30099, True)**]

From comparing the fuzzer settings to the output results, it can be readily determined that the communicator architecture and data collection / assessment mechanism are working as expected. This data, thus, validates the design and functionality of the fuzzer system and shows its efficacy for use in this particular real-world application.

**7. Conclusions and Future Work**

This paper has described and conducted initial efficacy evaluation of a fuzzing system that collects data to support vulnerability detection and the population of a cybersecurity assessment system for critical infrastructure that cannot be readily directly penetration tested. The fuzzer described in this paper is designed to be highly adaptive and readily reconfigurable. It includes a core component that designs campaigns based on configuration file definitions. It then tasks individual fuzzing activities to mechanisms that change the configuration file and detect exposed vulnerabilities and the impact of configuration file settings changes.

The data collected by this fuzzer system supports gaining a full understanding of the impact of different configuration settings on the actual operations of computer and networking equipment. It can detect vulnerabilities as well as characterize the actual operations of particular settings and combinations of settings. This data can then be supplied to the cybersecurity analysis system to ascertain the actual impact of current configuration settings and configuration changes on the operations of the system being assessed. Thus, the impact of configuration is able to be taken into account when determining attack pathways through the network.

While the addition of this information does not make the assessment system as accurate as testing on the actual system or an identically configured simulator, it can increase the fidelity of the analysis somewhat. It allows the assessment system to better represent the real world operations of the system and, thus, detect potential vulnerabilities.

While the fuzzer adds to the capabilities of this system somewhat, there are a number of other areas where additional data could be helpful. Additional accuracy can be gained through the collection of data regarding the operations of networking equipment, beyond just the impact of configuration settings, and through the collection of similar higher-fidelity data regarding the operations of computer systems. These will serve as important topics for potential future work.

**Appendix A. Parameter Generation Settings**

Name: start_systemctl_service
Type: bool
Default: true
Summary: When this configuration parameter is sent to the communicator it enables the systemd service for Apache2 when the value is set to true, and disables it when the value is set to false. This is used to verify the error outputs of several of the tests, but also shows that no ports are left open after the service has been quit.

Name: port

Type: number
Default: 80
Summary: This parameter changes which port the Apache2 HTTP server listens on and which port the default site configuration is enabled for. Port scans should only show this port when scanning for valid ports that the site could be running on.

Name: server_signature
Type: string
Default: "On"
Summary: When error pages are returned in response to an endpoint request, Apache2 – by default – also returns the server's version along with some other information. Changing this parameter changes what is returned in the version string or disables it completely.

Name: server_tokens
Type: string
Default: "OS"
Summary: HTTP responses from the Apache2 HTTP server include a header with the server's version and potentially the host's operating system. Changing this parameter changes what is included in this header, the minimum value that can be returned is "Apache". This is enabled by setting this parameter to "Min".

**Appendix B. Test Selection Settings**

Name: port scan
Summary: This is a basic TCP port scan of the target system. It is passed the port range defined by port_start and port_end. It attempts to connect to each port using a TCP connection and notes which ports it successfully connects to and which ports it fails to connect to.

Name: header version
Summary: This test makes a HEAD request to the HTTP server and notes what version is returned in the header "Server." The string that is in this header is defined by the config parameter ServerTokens (named server_token in the above configuration definition).

Name: page version
Summary: This test requests a page that will return a 404 server error and extracts the Apache server version that is displayed at the bottom of it. This value of this server version is determined by the config parameter ServerSignature (named server_signature in the above configuration definition).

Name: CVE-2021-41773
Summary: This test tests if the specified target server is vulnerable to CVE-2021-41773, a path traversal vulnerability. It tests for this by calling Metasploit with a resource script that executes the module *auxiliary/scanner/http/apache_normalize_path.* Note that this module can test for both CVE-2021-41773 and CVE-2021-42013.

Name: CVE-2021-42013
Summary: This test tests if the specified target server is vulnerable to CVE-2021-42013, a remote command execution vulnerability. It tests for this by calling the Metasploit with a resource script that executes the module *auxiliary/scanner/http/apache_normalize_path.* Note that this module can test for

both CVE-2021-41773 and CVE-2021-42013.

**Acknowledgements**

This work has been funded by the U.S. Missile Defense Agency (contract # HQ0860-22-C-6003).